\documentclass[12pt]{article}
\usepackage{graphicx}
\usepackage{epsfig}
\begin{document}
\begin{center}
{\bf FEYNMAN SCALING VIOLATION \\
ON BARYON SPECTRA IN $pp$ COLLISIONS \\
AT LHC AND COSMIC RAY ENERGIES}

\vspace{.2cm}

G. H. Arakelyan$^*$, C. Merino$^{**}$, C. Pajares$^{**}$,
and Yu. M. Shabelski$^{***}$ \\

\vspace{.5cm}
$^*$ A. I. Alikhanyan National Science Laboratory (Yerevan Physics
Institute)\\
Yerevan, Armenia\\
E-mail: argev@mail.yerphi.am

\vspace{.2cm}

$^{**}$ Departamento de F\'\i sica de Part\'\i culas and
\\
Instituto Galego de F\'\i sica de Altas Enerx\'\i as (IGFAE)\\
Universidade de Santiago de Compostela\\
Santiago de Compostela, Galiza, Spain\\
E-mail: merino@fpaxp1.usc.es \\
E-mail: pajares@fpaxp1.usc.es \\

\vspace{.2cm}

$^{***}$ Petersburg Nuclear Physics Institute,\\
NRC Kurchatov Institute,\\
St.Petersburg, Russia\\
E-mail: shabelsk@thd.pnpi.spb.ru

\vspace{.5cm}

A b s t r a c t
\end{center}

A significant asymmetry in baryon/antibaryon yields in the central
region of high energy collisions is observed when the initial state
has non-zero baryon charge. This asymmetry is connected with the
possibility of baryon charge diffusion in rapidity space. Such a diffusion 
should decrease the baryon charge in the fragmentation region and translate 
into the corresponding decrease of the multiplicity of leading baryons. 
As a result, a new mechanism for Feynman scaling violation in the 
fragmentation region is obtained. Another numerically more significant 
reason for the Feynman scaling violation comes from the fact that the 
average number of cutted Pomerons increases with initial energy. We present 
the quantitative predictions of the Quark-Gluon String Model (QGSM) for the 
Feynman scaling violation at LHC energies and at even higher energies that 
can be important for cosmic ray physics.

\vskip 1.cm


\newpage

\section{Introduction}


The problem of Feynman scaling violation has evident both theoretical and
practical interest. In particular, this question is very
important \cite{Abr,Abb} for cosmic ray physics, where the difference
from the primary radiation to the events registrated on the ground or
mountain level is determined by the multiple interactions of
the so-called leading particles (mainly baryons) in the atmosphere.

Despite the lack of direct measurements of Feynman scaling violation for
secondary baryon spectra in nucleon-nucleon collisions at energies higher
than those of ISR, some experimental information from cosmic ray
experiments seems to confirm \cite{ELF,Ver,KKh} the presence of significant
Feynman scaling violation effects. Now the LHCf Collaboration has started the
search \cite{LHCf1,LHCf2} of Feynman scaling violation effects for the
spectra of photons ($\pi^0$) and neutrons in the fragmentation region
at LHC energies.

In principle, the violation of Feynman scaling in the fragmentation region
should exist due to the energy conservation, since the spectra of
charged particles increase in the central region. However, no quantitative
predictions can be made without some model of the particle production.

The Additive Quark Model \cite{ASS,ABS} predicts the violation of Feynman
scaling in the fragmentation region due to the increase of the 
interaction cross sections. However, to make a description of the energy 
dependences of the spectra as a function of $x_F$ additional assumptions 
and parameters are needed.

The QGSM~\cite{KTM,Kaid} allows the calculation 
of the spectra of secondaries at different initial energies in the whole 
$x_F$ region. The QGSM is based on Dual Topological Unitarization (DTU), 
Regge phenomenology, and nonperturbative notions of QCD. This model is 
successfully used for the description of multiple production processes in 
hadron-nucleon \cite{KaPi,Sh,Ans,ACKS}, hadron-nucleus \cite{KTMS,Sh1}, and
nucleus-nucleus \cite{JDDS} collisions. The quantitative predictions of 
the QGSM depend on several parameters which were fixed by comparison of the 
calculations to the experimental data obtained at fixed target energies. 
The first experimental data obtained at LHC \cite{MPRS,MPS} show that the 
model predictions are in reasonable agreement with the data.

We will consider the energy dependences of the spectra of secondary 
baryons in the projectile fragmentation region which we determine as 
the interval $0.05 < x_F < 0.8$. These values of $x_F$ are larger 
than the typical values for central production and smaller than
the values where triple-Reggeon diagrams dominate.

In the frame of QGSM several reasons for the Feynman scaling violation
in the fragmentation region exist~\cite{AMPSh11}. The first one is the increase of the
average number of exchanged Pomerons with the energy, which leads to the
corresponding increase of the yields of hadron secondaries in the central
region and to their decrease in the fragmentation region. This effect is present
even at asymptotically high energies. The preliminary estimation of this effect
was provided in~\cite{Sh2,Sh3}.

In the case of nuclear (air) targets, the growth of the $hN$ cross section
with energy leads to the increase of the average number of fast hadron
inelastic collisions inside the nucleus. Thus, the average number of Pomerons
is additionally increased, resulting in a stronger Feynman scaling
violation~\cite{Sh2,Sh3}.

In ~\cite{EKS} these predictions were taken into account to calculate
the penetration of fast hadrons into the atmosphere, leading to a better
description of the cosmic ray experimental data.

The differences in the yields of baryons and antibaryons produced in the
central (midrapidity) region of high energy $pp$ interactions
\cite{ACKS,MPRS,MPS,BS,AMPS,MRS1,AKMS} are significant. Evidently,
the appearance of the positive baryon charge in the central region of $pp$
collisions should be compensated by the decrease of the baryon multiplicities
in the fragmentation region that leads to an additional reason for Feynman
scaling violation. This effect has a preasymptotical behaviour and it is
saturated at very high energies (see section~4).
 
In the present paper we consider the effects of Feynman scaling violation,
i.e. the energy dependences of the spectra of secondary protons, neutrons, 
and $\Lambda$ produced in $pp$ collisions in the fragmentation region.

In our estimations the role of the nuclear factor for air nuclei should be
similar to that presented in~\cite{Sh2,Sh3}.
 
\section{Inclusive spectra of secondary hadrons \newline in the
Quark-Gluon String Model}

The QGSM \cite{KTM,Kaid} allows us to make 
quantitative predictions for different features of multiparticle production, 
in particular, for the inclusive spectra of different secondaries, both in the 
central and in fragmentation regions. In QGSM high energy hadron-nucleon 
collisions are considered as taking place via the exchange of one or several 
Pomerons, all elastic and inelastic processes resulting from cutting through 
or between Pomerons~\cite{AGK}. 

Each Pomeron corresponds to a cylindrical diagram (see Fig.~1a), and thus, when 
cutting one Pomeron, two showers of secondaries are produced as it is shown in 
Fig.~1b. The inclusive spectrum of a secondary hadron $h$ is then determined 
by the convolution of the diquark, valence quark, and sea quark distributions, 
$u(x,n)$, in the incident particles, with the fragmentation functions, $G^h(z)$, 
of quarks and diquarks into the secondary hadron $h$. These distributions, as 
well as the fragmentation functions, are constructed by using the Reggeon counting 
rules~\cite{Kai}. Both the diquark and the quark distribution functions depend 
on the number $n$ of cut Pomerons in the considered diagram. The details of 
the model are presented in references~\cite{KTM,Kaid,KaPi,Sh,ACKS}.

\begin{figure}[htb]
\centering
\vskip -.3cm
\includegraphics[width=.65\hsize]{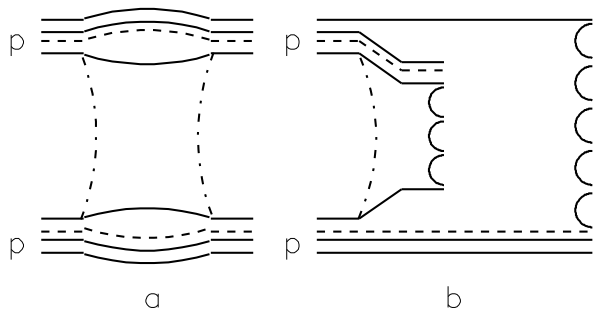}
\vskip -.5cm
\caption{\footnotesize
(a) Cylindrical diagram corresponding to the one--Pomeron exchange 
contribution to elastic $pp$ scattering, and (b) the cut of this diagram which 
determines the contribution to the inelastic $pp$ cross section. Quarks 
are shown by solid curves and the string junction by dashed curves.}
\end{figure}

For a nucleon target, the inclusive rapidity (y), or Feynman-$x$ ($x_F$), 
spectrum of a secondary hadron $h$ has the form~\cite{KTM}:
\begin{equation}
\frac{dn}{dy}\ = \  \frac1{\sigma_{inel}}\cdot \frac{d\sigma}{dy}\ =  
\frac{x_E}{\sigma_{inel}}\cdot \frac{d\sigma}{dx_F}\ = 
\sum_{n=1}^\infty w_n\cdot\phi_n^h (x) +  w_0\cdot\phi_D^h (x) \ ,
\end{equation}
where $x_{F}=2p_{\|}/\sqrt{s}$ is the Feynman variable, and
$x_{E}=2E/\sqrt{s}$, and the functions $\phi_{n}^{h}(x)$ determine the
contribution of the diagram with $n$ cut Pomerons and $w_n$ is the relative
weight of this diagram $\sum_{n=1}^\infty w_n = 1$. The last term in Eq.~(1)
accounts for the contribution of diffraction dissociation processes that are
determined by the cuts between Pomerons ($n=0$).

For $pp$ collisions
\begin{eqnarray}
\phi_{pp}^h(x) & = & f_{qq}^{h}(x_+,n)\cdot f_q^h(x_-,n) +
f_q^h(x_+,n)\cdot f_{qq}^h(x_-,n) + \nonumber\\
& + &  2(n-1)f_s^h(x_+,n)\cdot f_s^h(x_-,n)\ ,
\end{eqnarray}

\begin{equation}
x_{\pm} = \frac12\left[\sqrt{4m_T^2/s+x^2}\ \pm x\right] ,
\end{equation}
where $f_{qq}$, $f_q$, and $f_s$ correspond to the contributions of diquarks, 
valence quarks, and sea quarks, respectively.

These functions are determined by the convolution of the diquark and quark 
distributions with the fragmentation functions, e.g. for the quark one can 
write:
\begin{equation}
f_q^h(x_+,n)\ =\ \int\limits_{x_+}^1u_q(x_1,n)\cdot G_q^h(x_+/x_1) dx_1\ .
\end{equation}
The fragmentation functions $G^h(z)$ are independent on the number of cutted 
Pomerons $n$. On the contrary, the diquark and quark distributions $u(x,n)$ (which
are normalized to unity) become softer when $n$ increases. Thus, for example 
in \cite{KTM} it was assumed\footnote{There is some freedom \cite{KTMS} 
in how to account for this effect.} that the diquarks distributions depend on 
$n$ as:
\begin{equation}
u_{qq}(x) \sim (1-x)^{-\alpha_R+(n-1)} \;.           
\end{equation}

If the intercept of the Pomeron trajectory is larger than unity,
\begin{equation}
\alpha_P(0) = 1 + \Delta\;, \;\; \Delta > 0 \;,
\end{equation}
the average number of Pomerons which should be accounted for increases 
with energy. The probabilities for cutting different numbers of Pomerons, $n$,
$w_n$ in Eq.~(1), can be calculated in the quasieikonal approach~\cite{Kar3}. The
results of the calculation at four different energies, $\sqrt{s} = 17.3$ GeV, 200 GeV, 
8 TeV (the current LHC energy), and 100TeV (significant energy for the
Pierre Auger cosmic ray observatory, see, for example \cite{PAO}) are presented
in Fig.~2.
\begin{figure}[htb]
\centering
\vskip -.3cm
\includegraphics[width=.65\hsize]{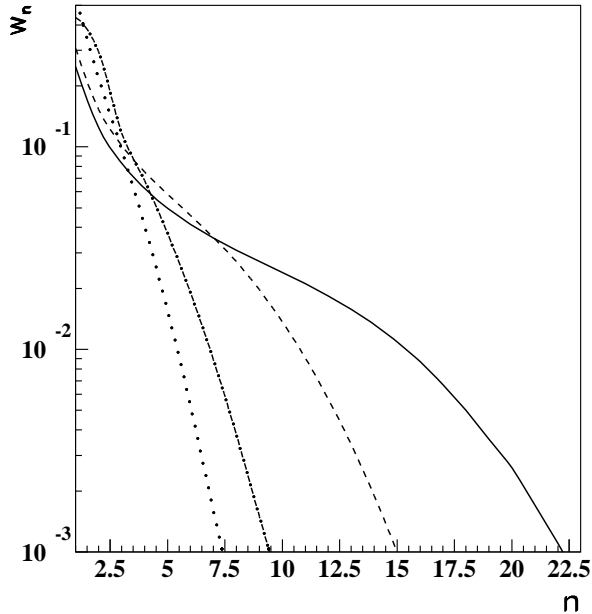}
\vskip -.5cm
\caption{\footnotesize
The calculated probabilities for cutting different number of Pomerons at
energies $\sqrt{s} = 17.3$ GeV (dotted curve), 200 GeV (dash-dotted curve),
8 TeV  (dashed curve), and 100 TeV (solid curve).}
\end{figure}

\section{Baryon/antibaryon asymmetry in the QGSM}

In the string models, baryons are considered as configurations
consisting of three connected strings (related to three valence quarks)
called string junction (SJ) \cite{Artru,IOT,RV,Khar}. Such a baryon
structure is supported by lattice calculations \cite{latt}. In the case of 
inclusive reactions the baryon number transfer to large rapidity distances 
in hadron-nucleon reactions can be explained \cite{ACKS,BS,AMS,Olga,MRS} 
by SJ diffusion.

The production of a baryon-antibaryon pair in the central region
occurs via $SJ$-$\overline{SJ}$ (SJ has upper color indices whereas
$\overline{SJ}$ has lower indices) pair production, which then combines 
with sea quarks and sea antiquarks into a $B\overline{B}$ pair \cite{RV,VGW}, 
as it is shown in Fig.~3a, the contributions of these processes to the 
inclusive spectra of secondary baryons being determined by Eq.~(2).
\begin{figure}[htb]
\centering
\includegraphics[width=.45\hsize]{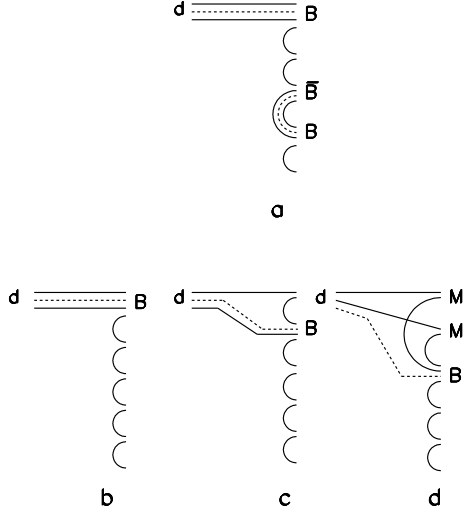}
\caption{\footnotesize
QGSM diagrams describing secondary baryon $B$ production by diquark $d$.
(a) Central production of $\overline{B}B$ pair. Single $B$ production
in the processes of diquark fragmentation: (b) initial SJ together with two
valence quarks and one sea quark, (c) initial SJ together with one valence
quark and two sea quarks, and (d) initial SJ together with three sea quarks.
Quarks are shown by solid curves and SJ by dashed curves.}
\end{figure}

In the processes with incident baryons, e.g. in $pp$ collisions, another
possibility to produce a secondary baryon in the central region exists.
This possibility is the diffusion in rapidity space of any SJ existing
in the initial state and it can lead to significant differences in the yields
of baryons and antibaryons in the midrapidity region even at high
energies~\cite{ACKS}. The most important experimental fact
in favour of this process is the rather large asymmetry in
$\Omega$ and $\overline{\Omega}$ baryon production in high energy $\pi^-p$
interactions~\cite{ait}.

The theoretical quantitative description of the baryon number transfer
via SJ mechanism was suggested in the 90's and used to predict~\cite{KP1}
the $p/\overline{p}$ asymmetry at HERA energies.

In order to obtain the net baryon charge we consider, following ref.\cite{ACKS} 
three different possibilities. The first one is the fragmentation
of the diquark giving rise to a leading baryon (Fig.~3b). A second possibility
is to produce a leading meson in the first break-up of the string and a baryon
in a subsequent break-up (Fig.~3c). In these two first cases the baryon
number transfer is possible only for short distances in rapidity. In the
third case, shown in Fig.~3d, both initial valence quarks recombine with
sea antiquarks into mesons $M$ while a secondary baryon is formed by the
SJ together with three sea quarks.

The fragmentation functions for the secondary baryon $B$ production
corresponding to the three processes shown in Figs.~3b, 3c, and 3d, can
be written as follows~\cite{ACKS}:
\begin{eqnarray}
G^B_{qq}(z) &=& a_N\cdot v^B_{qq} \cdot z^{2.5} \;, \\
G^B_{qs}(z) &=& a_N\cdot v^B_{qs} \cdot z^2\cdot (1-z) \;, \\
G^B_{ss}(z) &=& a_N\cdot\varepsilon\cdot v^B_{ss} \cdot z^{1 - \alpha_{SJ}}
\cdot (1-z)^2 \;,
\end{eqnarray}
where $a_N$ is the normalization parameter, and $v^B_{qq}$, $v^B_{qs}$,
$v^B_{ss}$ are the relative probabilities for different baryons production
that can be found by simple quark combinatorics \cite{AnSh,CS}. Their
numerical values for different secondary baryons were presented in
\cite{AKMS}.

The first two processes shown in Figs.~3b and 3c, Eqs.~(7) and (8),
determine the spectra of leading baryons in the fragmentation region. The
third contribution shown in Fig.~3d, Eq.~(9), is essential if the value of
the intercept of the SJ exchange Regge-trajectory, $\alpha_{SJ}$, is not
too small. In QGSM the weight of this third contribution is determined by
the coefficient $\varepsilon$ which fixes the small probability for such
a baryon number transfer to occur.

In the case of $pp$ collisions the most sensitive ratios to the values of
parameters $\alpha_{SJ}$ and $\varepsilon$ are the ratios of $\bar{B}/B$
in the central region. If the initial energy is high enough, one can neglect the 
contributions of the processes of figs.~3b and 3c, so the $\bar{B}/B$ ratio is 
determined by the contributions of figs.~3a and 3d. 

The spectra of antibaryons are described by QGSM \cite{KaPi,Sh,ACKS}
with reasonable accuracy, so the values of the parameters which determine the 
contributions of Fig.~3d can be extracted from the experimental data.

The energy dependence of $\bar{p}/p$ produced in $pp$ collisions in
midrapidity region ($y_{cm} \sim 0$) is shown in Fig.~4.
\begin{figure}[htb]
\vskip -.7cm
\centering
\includegraphics[width=.6\hsize]{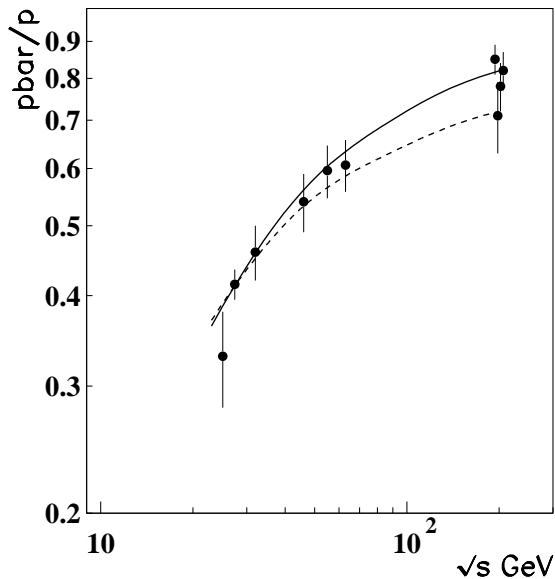}
\vskip -.5cm
\caption{\footnotesize
The QGSM description of the energy dependence of the $\bar{p}/p$ in the
midrapidity region. Solid curve correspond to the value $\alpha_{SJ} = 0.5$
and dashed curve to the value $\alpha_{SJ} = 0.9$.}
\end{figure}
The theoretical curve has been normalized to the experimental point at 
$\sqrt{s}$ = 27.5 GeV, where the error bar is minimal and the $\chi^2$
analyses gives \cite{MRS1}:
\begin{equation}
\alpha_{SJ} = 0.5 \pm 0.1 \;, {\rm with}\;\;  \varepsilon = 0.0757 \;.
\end{equation}

Unfortunately, four experimental points at RHIC energy $\sqrt{s}$ = 200 GeV are
in evident disagreement with each other, the calculation performed with values
of the parameters $\alpha_{SJ} = 0.9$ and $\varepsilon = 0.024$ being in agreement
with the lowest RHIC point (see dashed curve in Fig.~4).

The data by the ALICE Collaboration \cite{ALICE} for $\bar{p}/p$ ratios in 
$pp$ collisions at $\sqrt{s} = 900$ GeV and 7 TeV in midrapidity region are 
presented in Table 1, together with the QGSM predictions. These data confirm 
the asymmetry in the ratio $\bar{p}/p$ and they are in agreement with the QGSM 
predictions for the value $\alpha_{SJ} = 0.5$.
\begin{center}
\begin{tabular}{|c||r|r|} \hline
SJ exchange & $\sqrt{s} = 900$ GeV & $\sqrt{s} = 7$ TeV \\ \hline
$\alpha_{SJ} = 0.9$ & 0.89\hspace{1.cm} & 0.95\hspace{1.cm} \\
$\alpha_{SJ} = 0.5$ & 0.95\hspace{1.cm} & 0.99\hspace{1.cm} \\
$\varepsilon = 0$\hspace{0.2cm}  & 0.98\hspace{1.cm} & 1.\hspace{1.4cm}   
\\ \hline
ALICE & $0.957$ \hspace{1.cm} & $0.991$ \hspace{1.cm} \\
Collaboration & $\pm 0.006 \pm 0.014$ & $\pm 0.005 \pm 0.014$ 
\\ \hline
\end{tabular}
\end{center}
Table 1. {\footnotesize The QGSM predictions for $\bar{p}/p$ in $pp$
collisions at LHC energies and the corresponding data by the ALICE Collaboration.
The value $\varepsilon = 0$ corresponds to the case without C-negative exchange.}
\vskip 5pt

The LHCb Collaboration measured the ratios of $\bar{\Lambda}$ to $\Lambda$ 
in the rapidity interval $2<y<4$ at $\sqrt{s} = 900$ GeV and 7 TeV
\cite{LHCb}. These preliminary data are also incompatible \cite{MPS3}
with the QGSM calculation without SJ contribution ($\varepsilon = 0$). Though
the errorbars are too large to make any conclusion, the QGSM calculations with the
value $\alpha_{SJ} = 0.9$ seem to be in a slightly better agreement with the data 
than the calculations with $\alpha_{SJ} = 0.5$.

So, we can conclude that the transfer of baryon charge in large rapidity
distances occurs up to the LHC energies, and it can probably be described 
by the QGSM by taking the values of the parameters $\alpha_{SJ}$
and $\varepsilon$ presented in Eq.~(10).

\section{Energy dependence of secondary proton spectra in 
the fragmentation region}

As it was mentioned above, the inclusive spectra of secondary net baryons 
$B$ produced in the processes of Figs.~3b, 3c, and 3d are determined by the
convolution of the diquark distribution $u(x,n)$ in the incident particles 
with the fragmentation functions $G^h(z)$ presented in Eqs.~(7)$-$(9). The
baryon charge of all secondary particles is determined by the integral
over these spectra and it should be excatly equal to two (i.e. to the 
baryon charge of the initial $pp$ state). In the case the process in
Fig.~3d is absent, the integral over the spectra of secondary net baryons 
totally saturates at the distance of several units of rapidity from the
projectile. Thus, in the absence of the process in Fig.~3d the Feynman
scaling violation in the fragmentation region can only be due to effects
connected with the increase of the average number of exchanged Pomerons.

The SJ contribution to the inclusive cross section of secondary baryon 
production (Fig.~3d) at large rapidity distance $\Delta y$ from
the incident nucleon can be estimated as
\begin{equation}
\frac{1}{\sigma}\frac{\sigma^B}{dy} \sim a_B\cdot\varepsilon\cdot
e^{(1 - \alpha_{SJ})\cdot\Delta y} \;,\;\;
a_B = a_N\cdot v^B_{ss} \;.
\end{equation}
At asymptotically high energies, the baryon charge transferred to large
rapidity distances can be determined by integration of Eq.~(11), and it turns out
to be of the order of
\begin{equation}
\langle n_B(s \to \infty) \rangle_{SJ} 
\sim a_B\cdot\frac{\varepsilon}{(1 - \alpha_{SJ})} \; ,
\end{equation}
only the left part of the initial baryon charge being available for
the production of the leading baryons.

The only free parameter in eqs.~(7)$-$(9) is the normalization $a_N$, that
should be modified at asymptotically high energies in the presence of the SJ
mechanism for the baryon charge as
\begin{equation}
\tilde{a}_N(s \to \infty) = a_N\cdot\frac{\langle n_B \rangle_{\varepsilon=0}}
{\langle n_B \rangle_{\varepsilon=0} + 
\langle n_B(s \to \infty)  \rangle_{SJ}} \; ,
\end{equation}
to guarantee the conservation of the baryon charge. One can see that
$\langle n_B(s \to \infty) \rangle_{SJ}$ is finite if $\alpha_{SJ} < 1$
(see Eq~(12)), so the Feynman scaling violation due to the discussed effects 
has a preasymptotical behaviour.

To obtain the QGSM predictions for the spectra of leading baryons at finite
energy $s$ we have to calculate the value of $\langle n_B(s) \rangle_{SJ}$ at
this energy for the renormalized value of $\tilde{a}_N(s)$ in Eq.~(13). This 
can be provided by the numerical integration of the convolution of the diquark 
distribution $u(x,n)$ in the incident protons with the fragmentation functions 
$G^h(z)$. By this way we account for the rather complicate shape of 
$(1/\sigma)\cdot d\sigma^B/dy$ at small $\Delta y$.

Though currently the value of $\alpha_{SJ} = 0.5$ seems more
plausible~\cite{ALICE}, the value of $\alpha_{SJ} = 0.9$ can not be
excluded~\cite{LHCb,Conf}. Thus, in this paper we present the calculation
obtained with these two values of $\alpha_{SJ}$,
and also without any SJ contribution ($\varepsilon = 0.$).

The results of the calculations of the secondary proton spectra produced in
$pp$ collisions are presented in Fig.~5, together with experimental data~\cite{na49p}
by the NA49 Collaboration on proton spectra in $pp$ collisions at 158 GeV/c beam momentum
($\sqrt(s)=17.3$ Gev) and the data \cite{Bren} at 100 GeV/c 
($\sqrt{s}$=14 GeV) and 175 GeV/c ($\sqrt{s}$=19 GeV).
\begin{figure}[htb]
\centering
\vskip -.6cm
\includegraphics[width=.49\hsize]{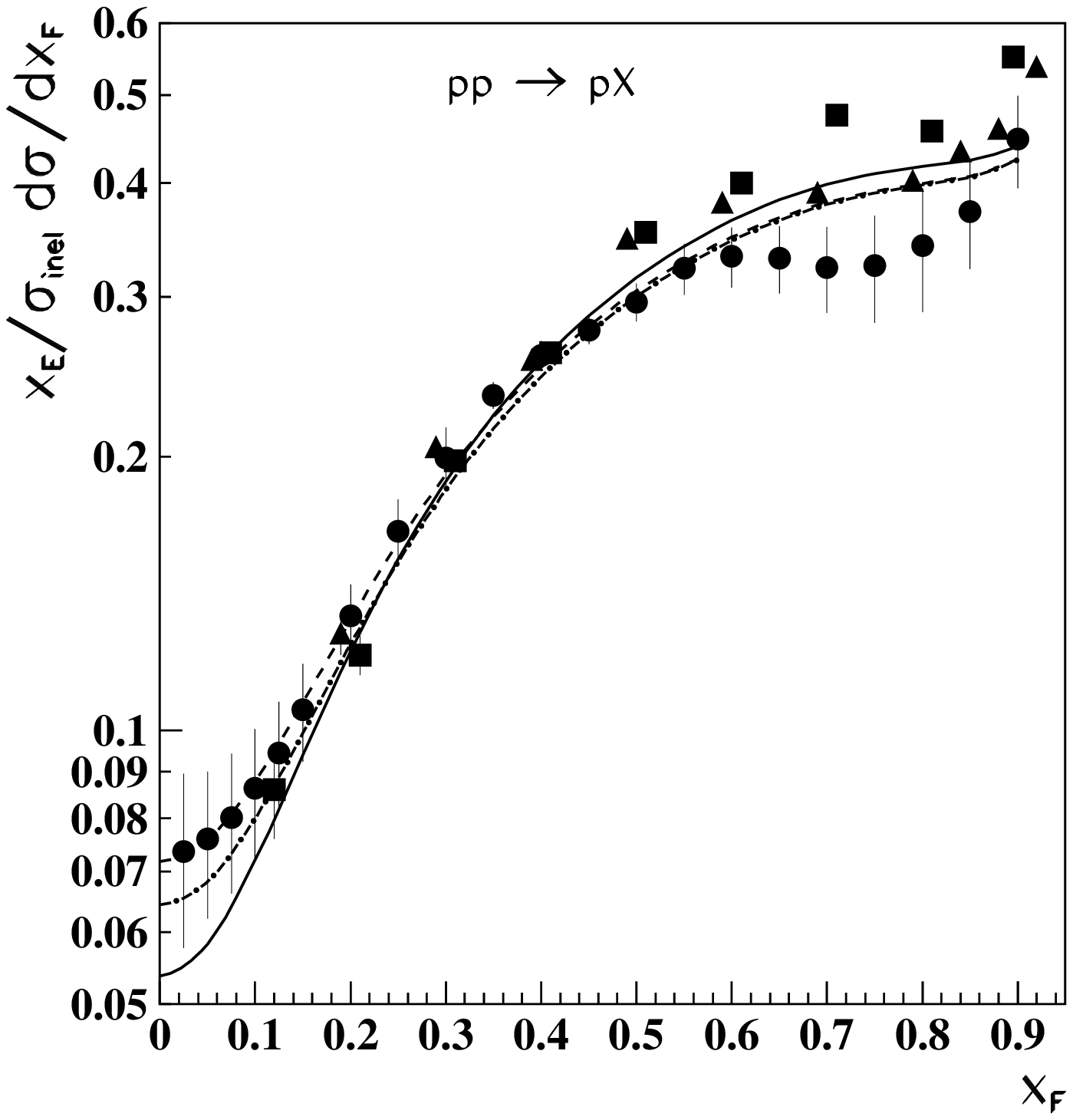}
\includegraphics[width=.49\hsize]{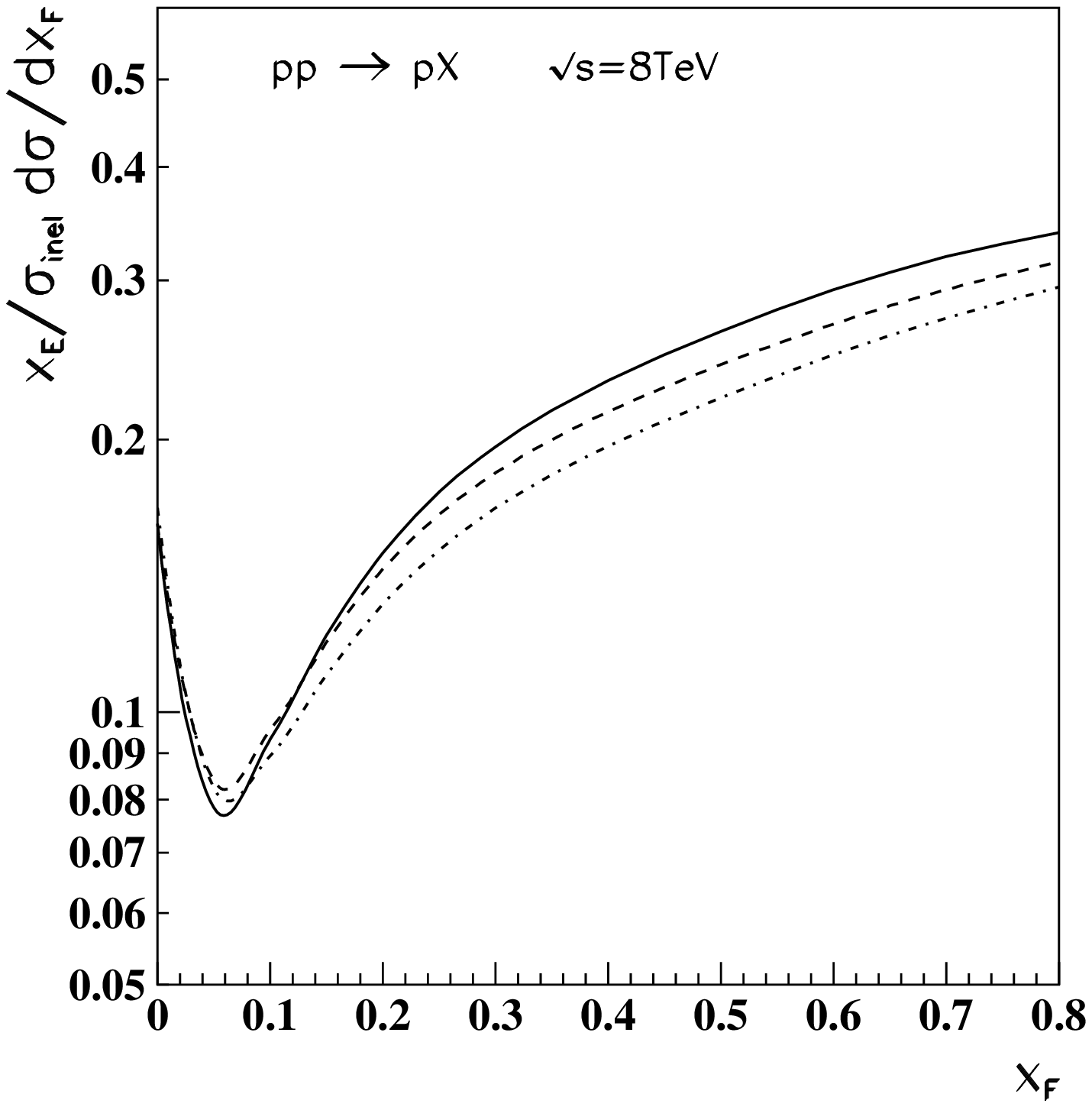}
\vskip -.4cm
\caption{\footnotesize 
The QGSM predictions for the spectra of secondary protons produced in $pp$ 
collisions at $\sqrt{s} = 17$ GeV (left) and 8 TeV (right), compared to the
corresponding experimental data by the NA49 collaboration~$\cite{na49p}$ at
$\sqrt{s} = 17$ GeV (points) and data \cite{Bren} at $\sqrt{s} = 14$ 
(triangles) and 19 (squares) GeV. 
Calculations without SJ contribution are shown by solid curves, results for 
$\alpha_{SJ} = 0.5$ by dashed curves, and for $\alpha_{SJ} = 0.9$ by 
dash-dotted curves.}
\end{figure}

As it is seen on Fig.~5, the comparison of the QGSM calculations to the experimental
data in ref.~$\cite{na49p,Bren}$ is rather good.The data of
these two experimental groups are in good agreement at $x_{\rm F} < 0,6$ and in
some disagreement at larger $x_{\rm F}$.  
The calculated spectra of secondary protons produced at  $\sqrt{s}$=17 GeV 
and 8 TeV are significantly different. 
The calculated spectrum at $\sqrt{s} = 17$ GeV
increases with $x_F$ rather monotonically, in agreement with the existing experimental
data. This spectrum only weakly depends on the SJ 
contributions, the reason being that the value of $\langle n_B(s) \rangle_{SJ}$ 
at this energy is rather small. With the increase of the energy until 
$\sqrt{s} = 8$ TeV the rapidity region accessible for the baryon number 
transfer increases, increasing also the value of $\langle n_B(s) \rangle_{SJ}$ 
and making the SJ effects more visible.

The peak appearing at $\sqrt{s} = 8$ TeV and $x_F < 0.06$ is connected with the
protons produced together with antiprotons via the mechanism shown in  Fig.~3a.

The proton spectra shown in Fig.~5 at both (low and high) energies depend on the 
values of
several QGSM parameters. These considered values were determined from the comparison
of the model calculations to the experimental data, but since there is some 
small disagreement between the calculations and the data.The main part of the
uncertainty being connected with the normalization), the values of parameters,
as well as the absolute values of the calculated spectra, can be known on the accuracy
of the order of, say, 20\%. 
Consequently, the difference between the calculations with
and without SJ contribution at the same energy can be estimated on the level of the 20\%
of this difference, in agreement with Eq.~(10).  
  
Similar situation appears when we consider the ratios of the spectra of
the same secondary $h$ at different energies: 
\begin{equation}
R_h(\frac{\sqrt{s_1}}{\sqrt{s_2}}) = \left [\frac{x_E}{\sigma_{inel}}\cdot 
\frac{d\sigma}{dx_F} \right ]_{s_1} \left / \right. 
\left [\frac{x_E}{\sigma_{inel}}\cdot \frac{d\sigma}{dx_F}\right ]_{s_2} \;.
\end{equation}
In Fig.~6 the QGSM predictions for the ratio $R_p(8 {\rm TeV} / 17 {\rm GeV})$ are 
presented.  
Here again the normalization uncertainties are canceled and the accuracy of the model 
predictions is good enough.
\begin{figure}[htb]
\centering
\vskip -.4cm
\includegraphics[width=.62\hsize]{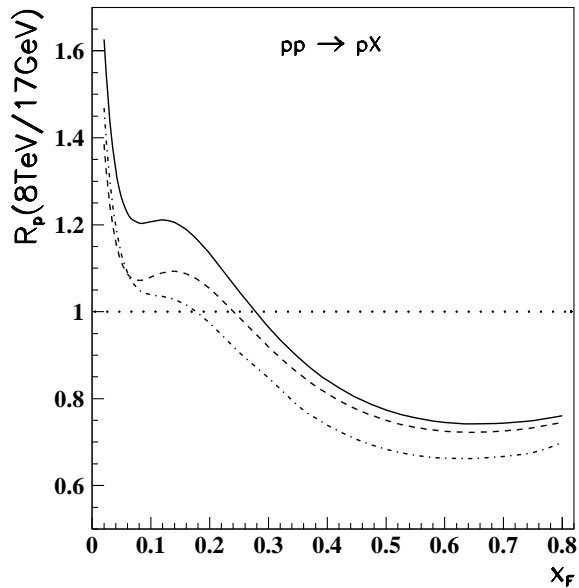}
\vskip -.8cm
\caption{\footnotesize 
The QGSM predictions for the  ratios of the spectra of secondary protons 
produced in $pp$ collisions at $\sqrt{s} = 17$ GeV and 8 TeV. Calculations 
without SJ contribution are shown by solid curves, results for 
$\alpha_{SJ} = 0.5$ by dashed curves, and for $\alpha_{SJ} = 0.9$ by dash-dotted 
curves.}
\end{figure}

When the initial energy increases the spectra of secondary protons with
$x_F > 0.3$ decrease, mostly due to the increase of the average number of 
cut Pomerons (see Fig.~2). The same effect increases the spectra at $x_F < 0.3$.  
Some additional contribution to these Feynman scaling violations comes from 
the SJ effects which transfer the baryon charge to the low-$x_F$ region 
resulting in the renormalization of the parameter $a_N$ (see Eq.~(13)).   

It is also interesting to consider the spectra of secondary protons  at fixed
points of $x_F$ as functions of initial energy. Here again, the absolute
normalization of the curve contains some uncertainties, but their relative
change with the energy and their dependence on the SJ effects can be calculated 
with reasonable accuracy. The results of such calculations for secondary 
protons with $x_F$ = 0.7, 0.5, 0.2, and 0.05 produced in $pp$ collisions at 
different energies are shown in Fig.~7.
\begin{figure}[htb]
\centering
\vskip -.4cm
\includegraphics[width=.42\hsize]{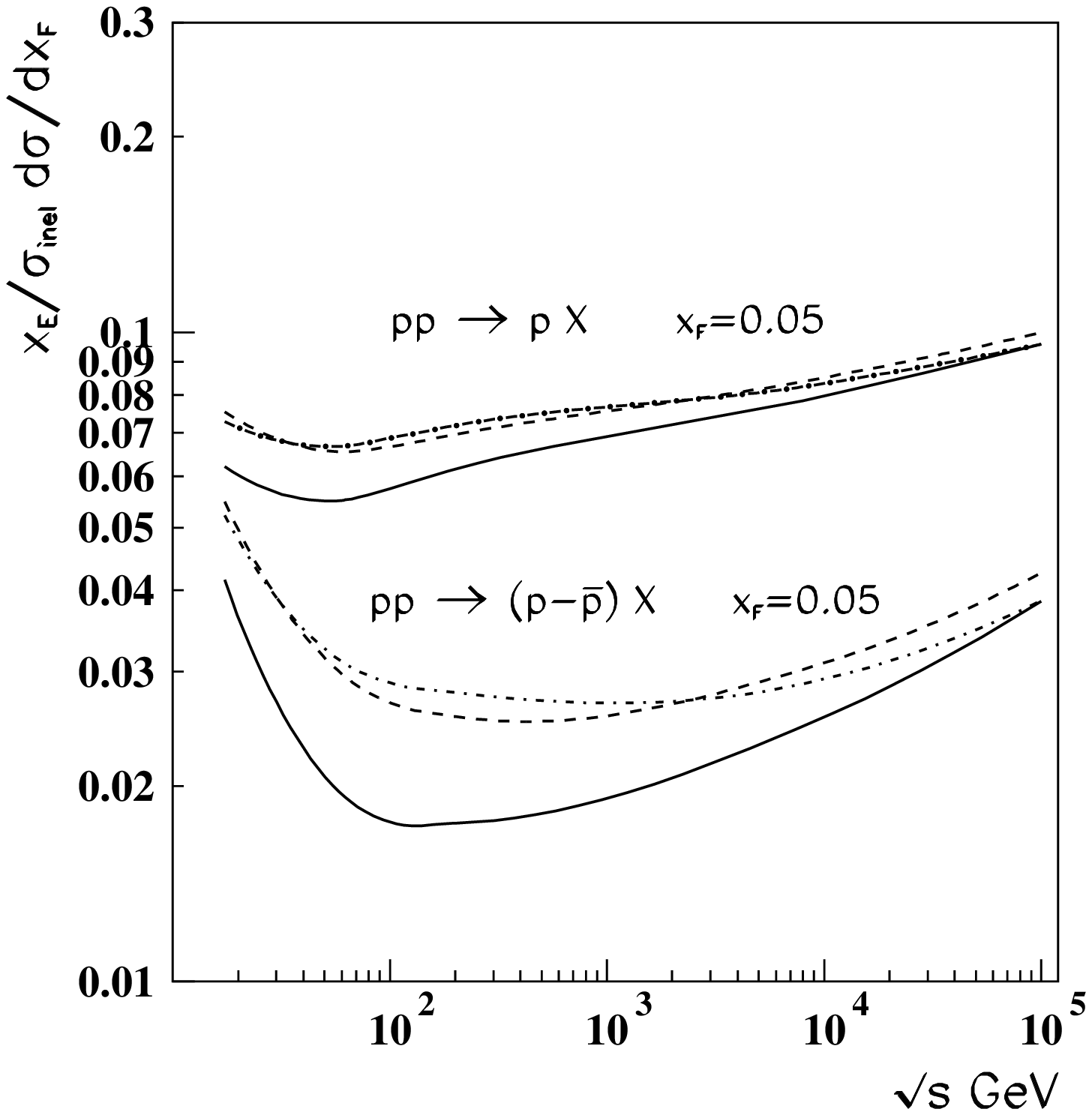}
\includegraphics[width=.42\hsize]{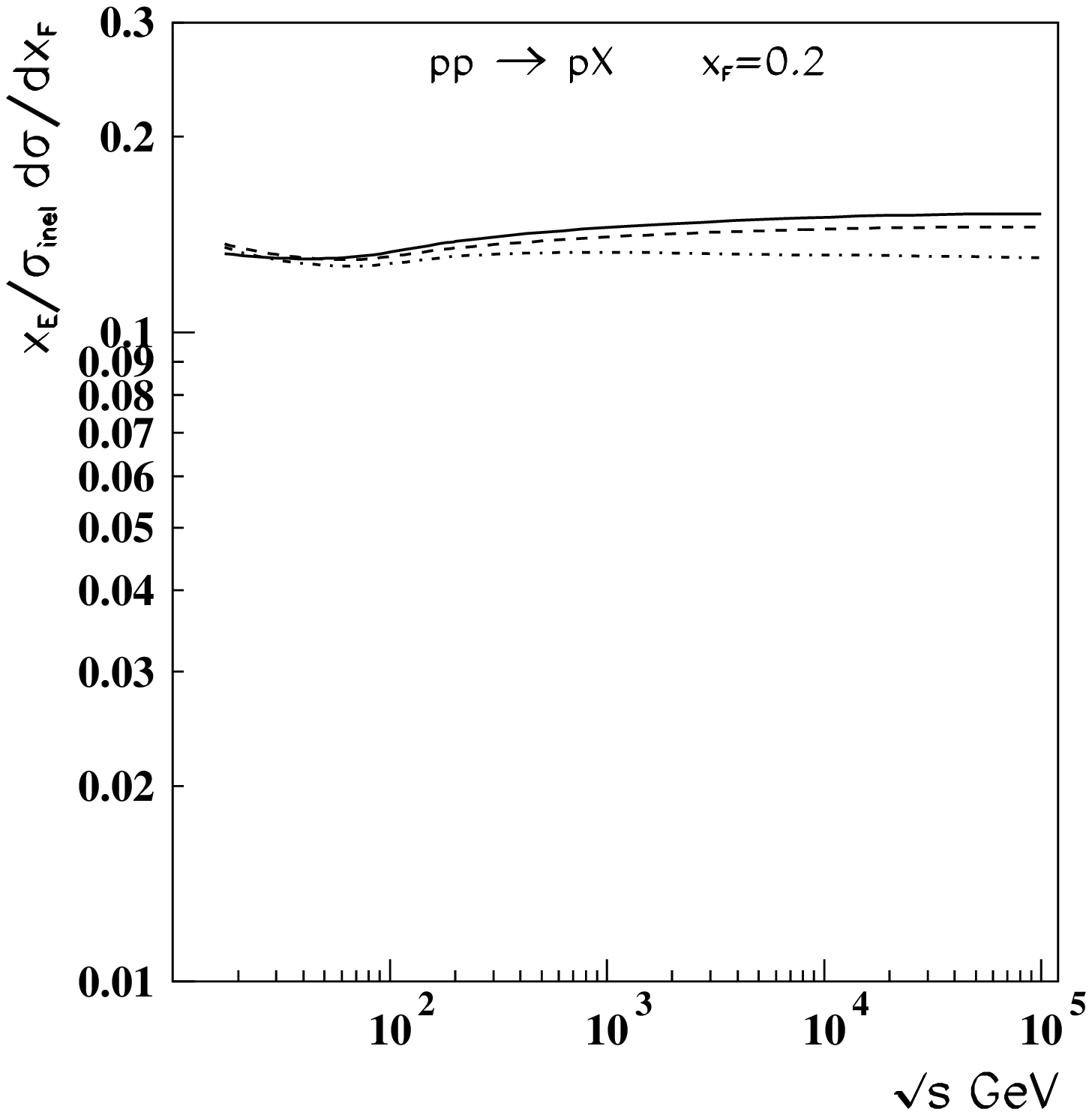}
\vskip -.4cm
\includegraphics[width=.42\hsize]{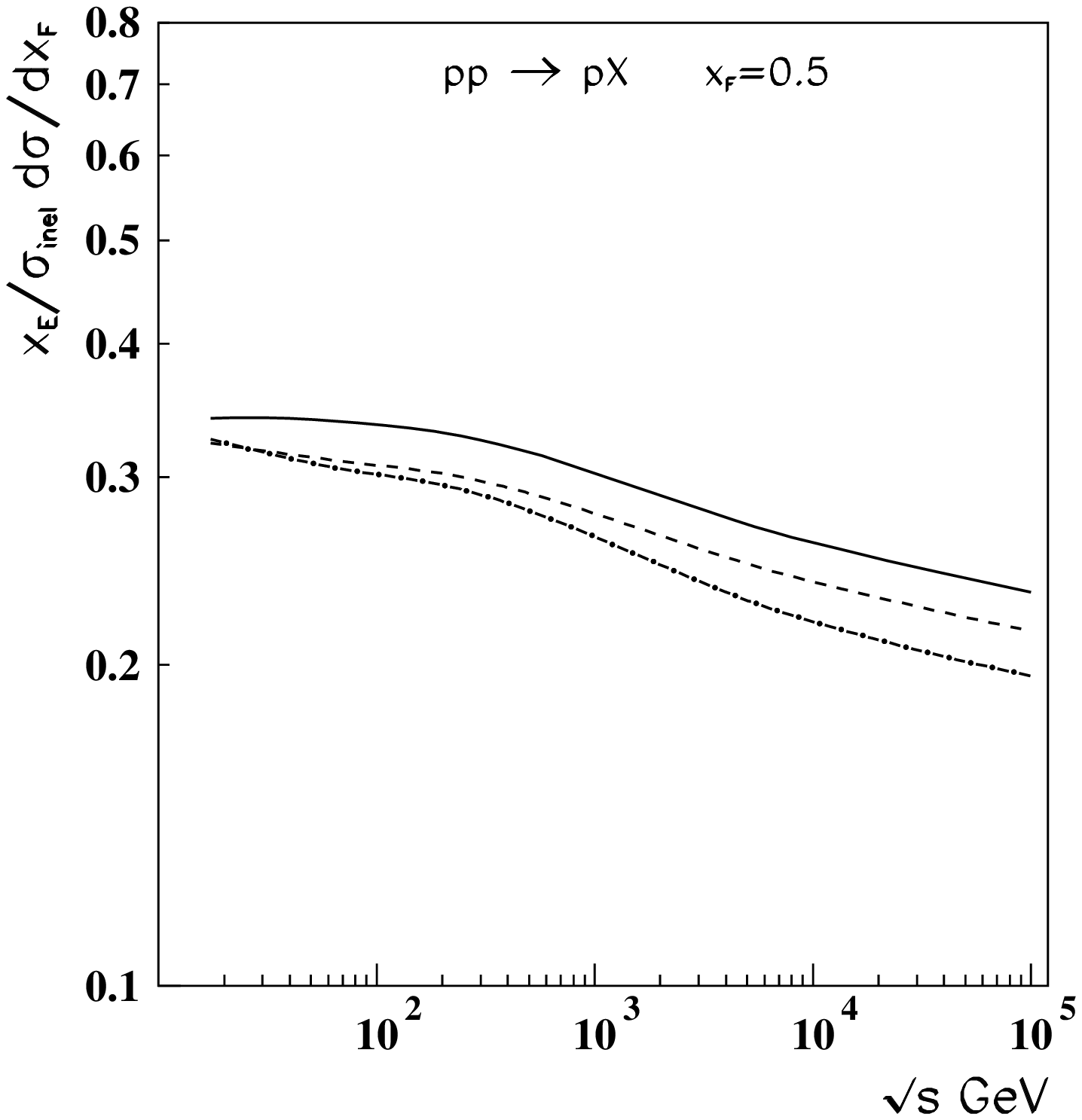}
\includegraphics[width=.42\hsize]{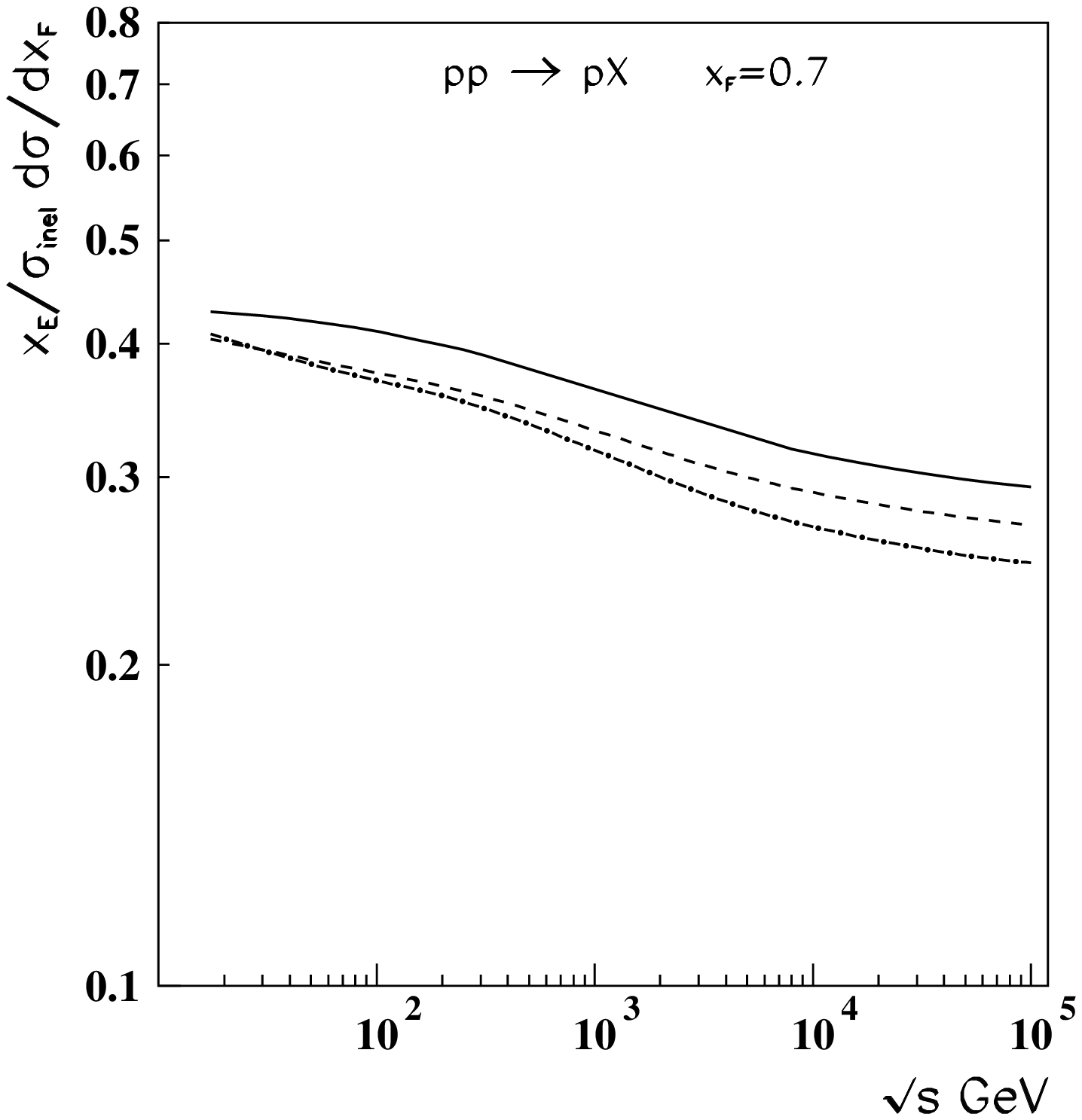}
\vskip -.4cm
\caption{\footnotesize
The QGSM predictions for the spectra of secondary protons as
functions of the energy at fixed values of $x_F$. The left top bold curves
show the spectra of net protons, i.e. the values of the $p-\overline{p}$
differences. For the four panel, solid curves correspond to the calculation
without SJ contribution, dashed curves to the value $\alpha_{SJ} = 0.5$,
and dash-dotted curves to the value $\alpha_{SJ} = 0.9$.}
\end{figure}

These spectra increase with energy at comparatively low $x_F$ = 0.05 and 0.2 
(except for the calculation with $\alpha_{SJ} = 0.9$ in the last case), and they 
decrease with energy at $x_F$ = 0.5 and 0.7. This change in the  energy 
dependence is explained by the growth of the average number of exchanging 
Pomerons (see Fig.~2). 

The important feature is that at low $x_F$ (about 0.05 and less) one should 
discriminate between the total yield of secondary protons and the yield of net 
protons, i.e. the values of the $p-\overline{p}$ differences, by comparing the 
upper and lower sets of curves in the left top panel of Fig.~7. The spectra of
antibaryons are not affected by SJ effects, so after subtraction of $\bar{p}$ spectra
the SJ effects are more visible in the spectra of net protons.

\section{Predictions for the spectra of neutrons and $\Lambda$ in 
$pp$ collisions}

The LHCf Collaboration plans to investigate the Feynman scaling violation for
neutral secondaries in the fragmentation region. In this perspective, the
QGSM predictions for the spectra of secondary neutrons and $\Lambda$-hyperons
produced in $pp$ collisions at $\sqrt{s} = 8$ TeV are presented in Fig.~8.
\begin{figure}[htb]
\centering
\vskip -.4cm
\includegraphics[width=.62\hsize]{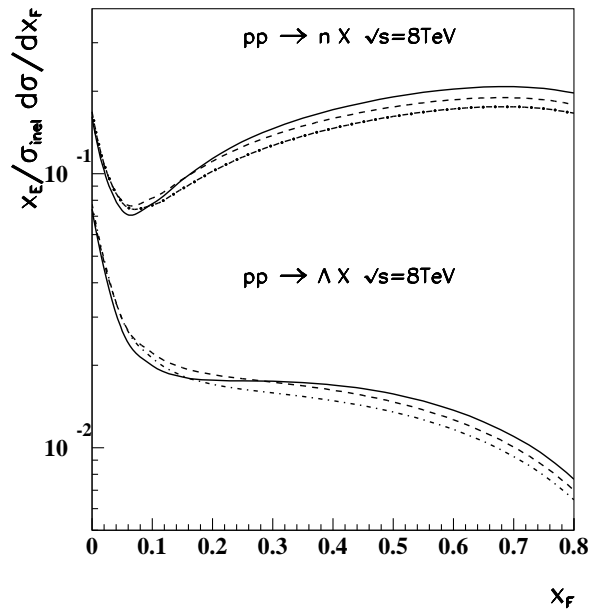}
\vskip -.5cm
\caption{\footnotesize 
The QGSM predictions for the spectra of secondary neutrons and of
$\Lambda$-hyperons produced in $pp$ collisions at $\sqrt{s} = 8$ TeV. 
Calculations without SJ contribution are shown by solid curves, results for 
$\alpha_{SJ} = 0.5$ by dashed curves, and for $\alpha_{SJ} = 0.9$ by dash-dotted 
curves.}
\end{figure}

The spectra of neutrons are similar to the spectra of secondary protons 
presented in Fig.~5 but the fragmentation maxima are not so stressed. The 
spectra of $\Lambda$-hyperons do not present such a maxima due to the additional 
suppression of fast strange particle production with respect to the non-strange 
secondaries, what leads to a faster decrease of $uu$ and $ud$ fragmentation
functions into $\Lambda$ at large $z$ in comparison with the fragmentation into 
secondary nucleon.

In both neutron and $\Lambda$ production cases the SJ effects lead to 
similar corrections of the spectra than for secondary protons.  

The ratios $R_n(8 {\rm TeV} / 17 {\rm GeV})$ and  
$R_{\Lambda}(8 {\rm TeV} / 17 {\rm GeV})$ of the spectra of neutrons and of 
$\Lambda$-hyperons produced in $pp$ collisions, defined by Eq.~(14), are shown 
in Fig.~9.
\begin{figure}[htb]
\centering
\vskip -.4cm
\includegraphics[width=.49\hsize]{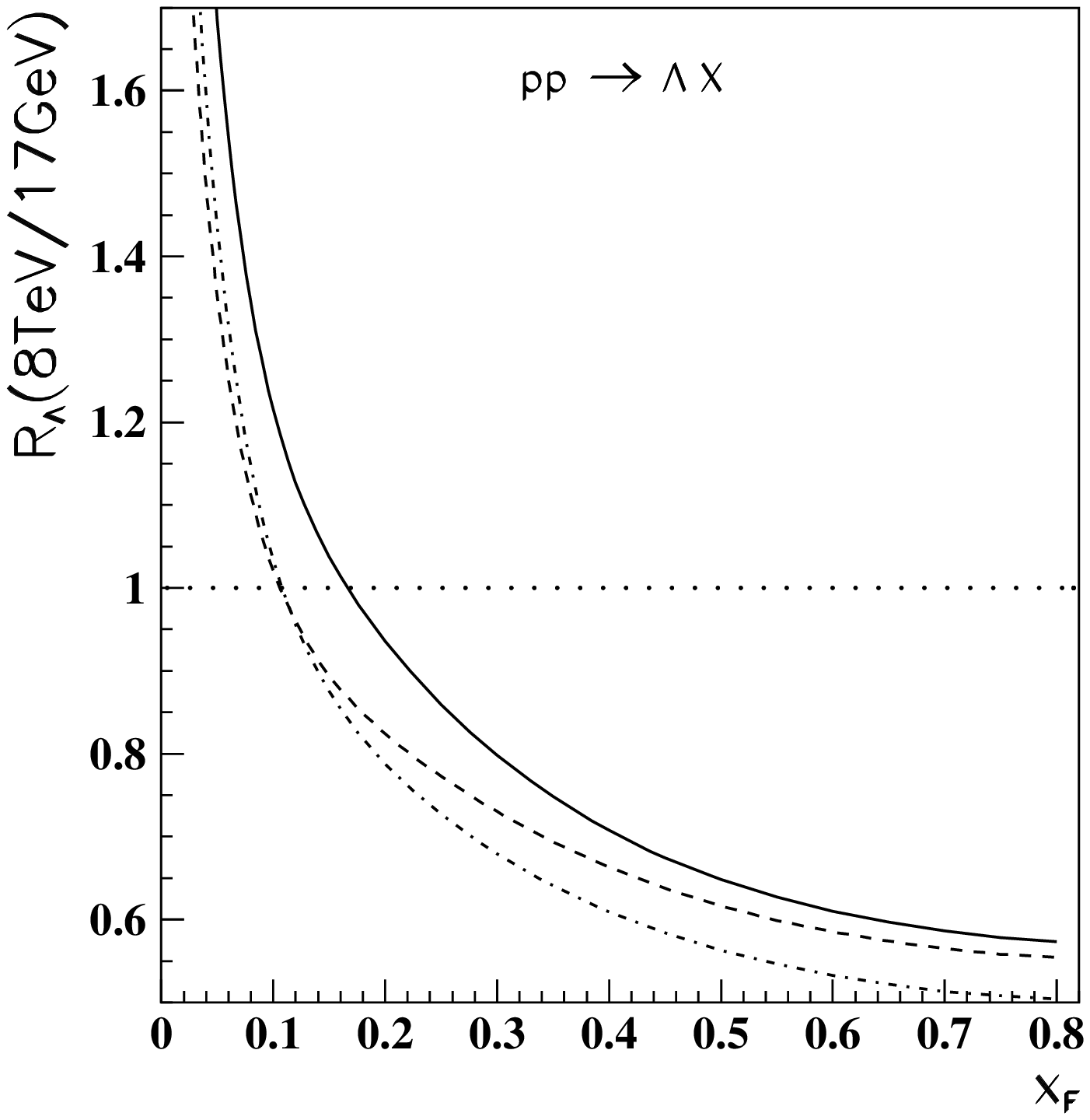}
\includegraphics[width=.49\hsize]{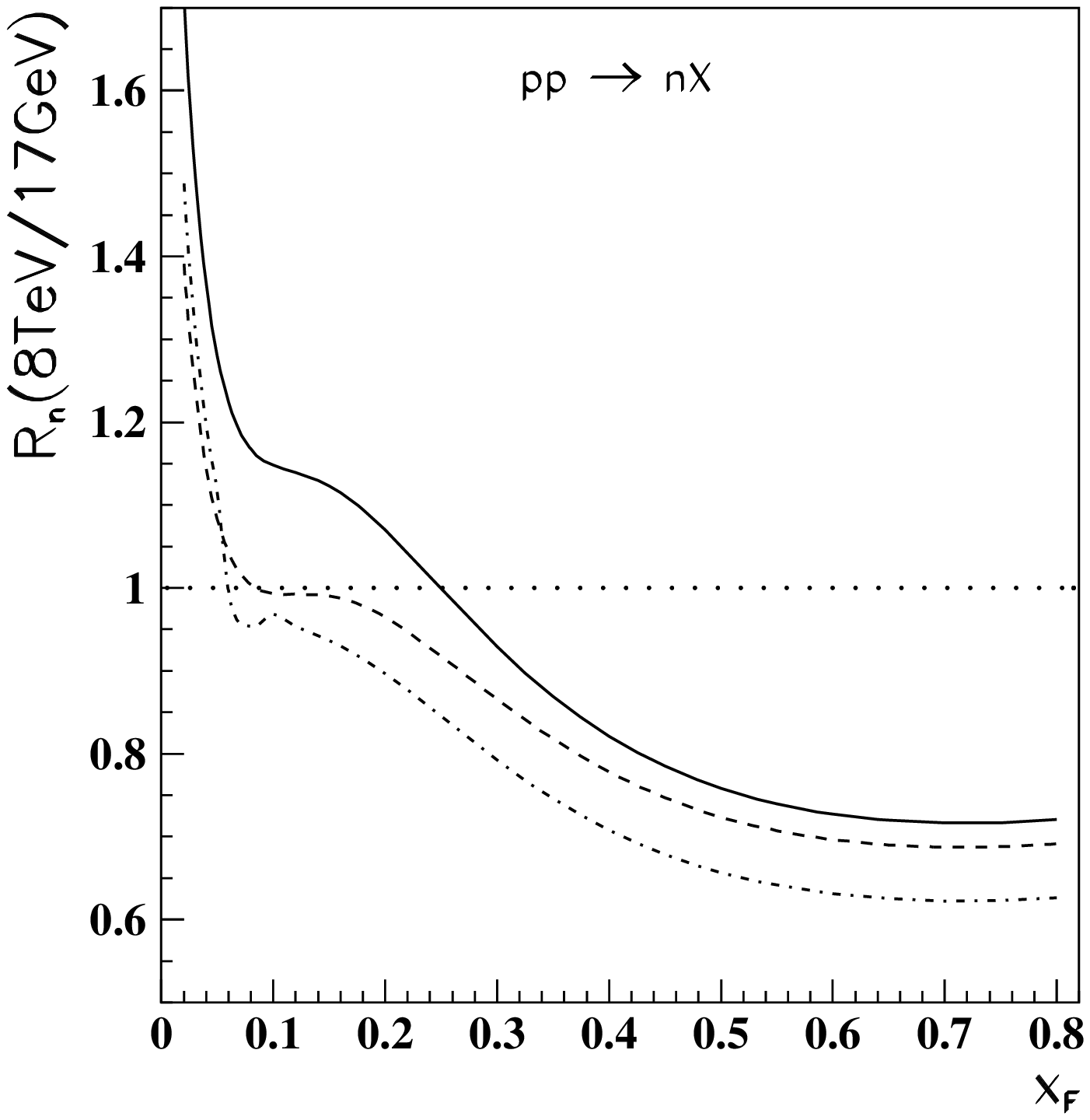}
\vskip -.4cm
\caption{\footnotesize
The QGSM predictions for the ratios of the spectra of secondary neutrons 
(left) and $\Lambda$-hyperons (right) as functions of $x_F$. The solid
curves correspond to the calculation without SJ contribution, dashed curves
to the value $\alpha_{SJ} = 0.5$, and dash-dotted curves to the value
$\alpha_{SJ} = 0.9$.}
\end{figure}
These ratios show the expected effects of the Feynman scaling 
violation in different $x_F$-regions. The ratios for secondary neutrons are
similar to the corresponding ratios for secondary protons presented in Fig.~6.
The ratios for secondary $\Lambda$ are slightly different due to the
difference in the fragmentation functions.

The energy dependences of the spectra of neutrons and $\Lambda$-hyperons at
$x_F$ = 0.05 and 0.5 are presented in Fig.~10.
\begin{figure}[htb]
\centering
\vskip -.4cm
\includegraphics[width=.49\hsize]{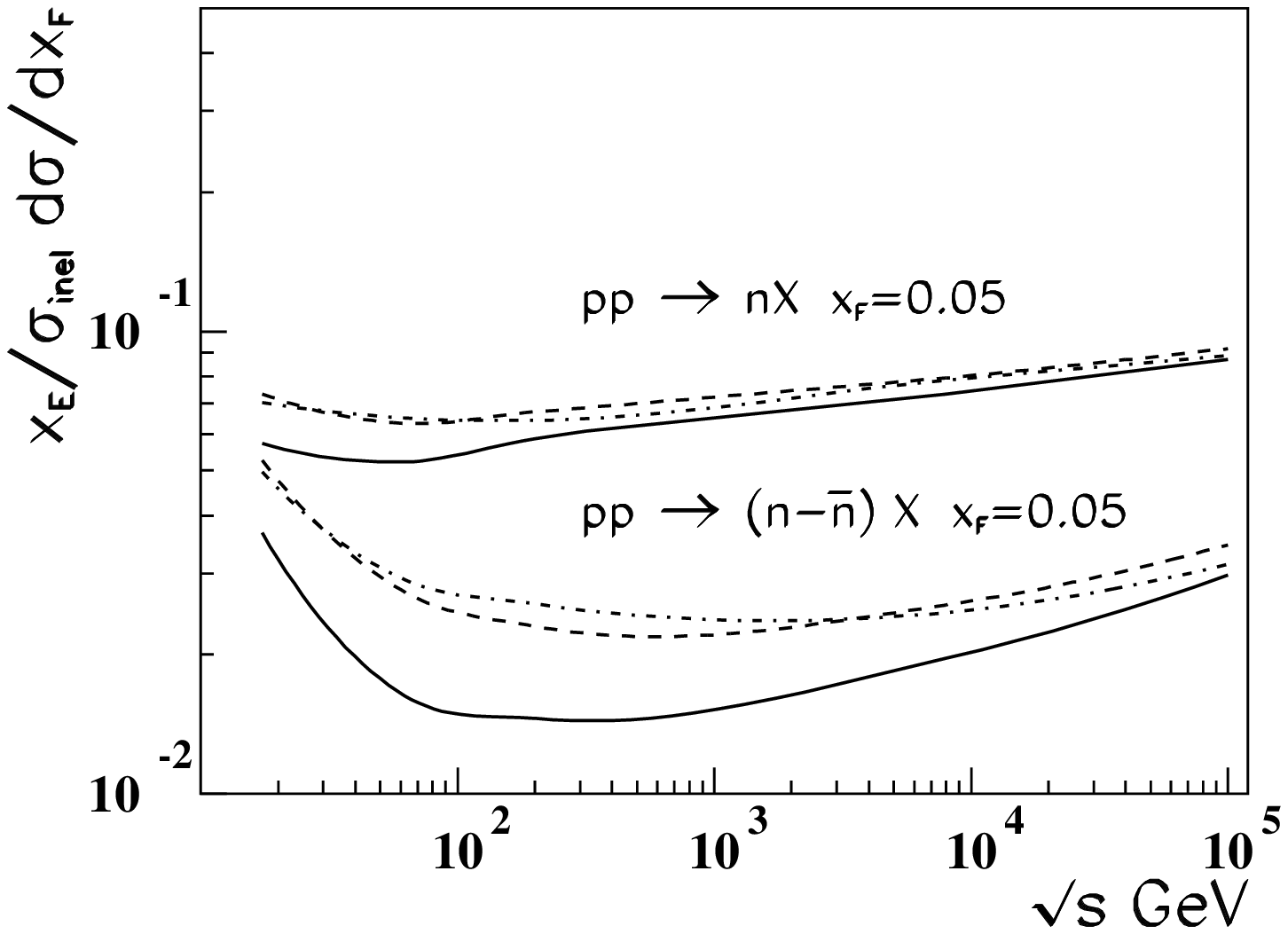}
\includegraphics[width=.49\hsize]{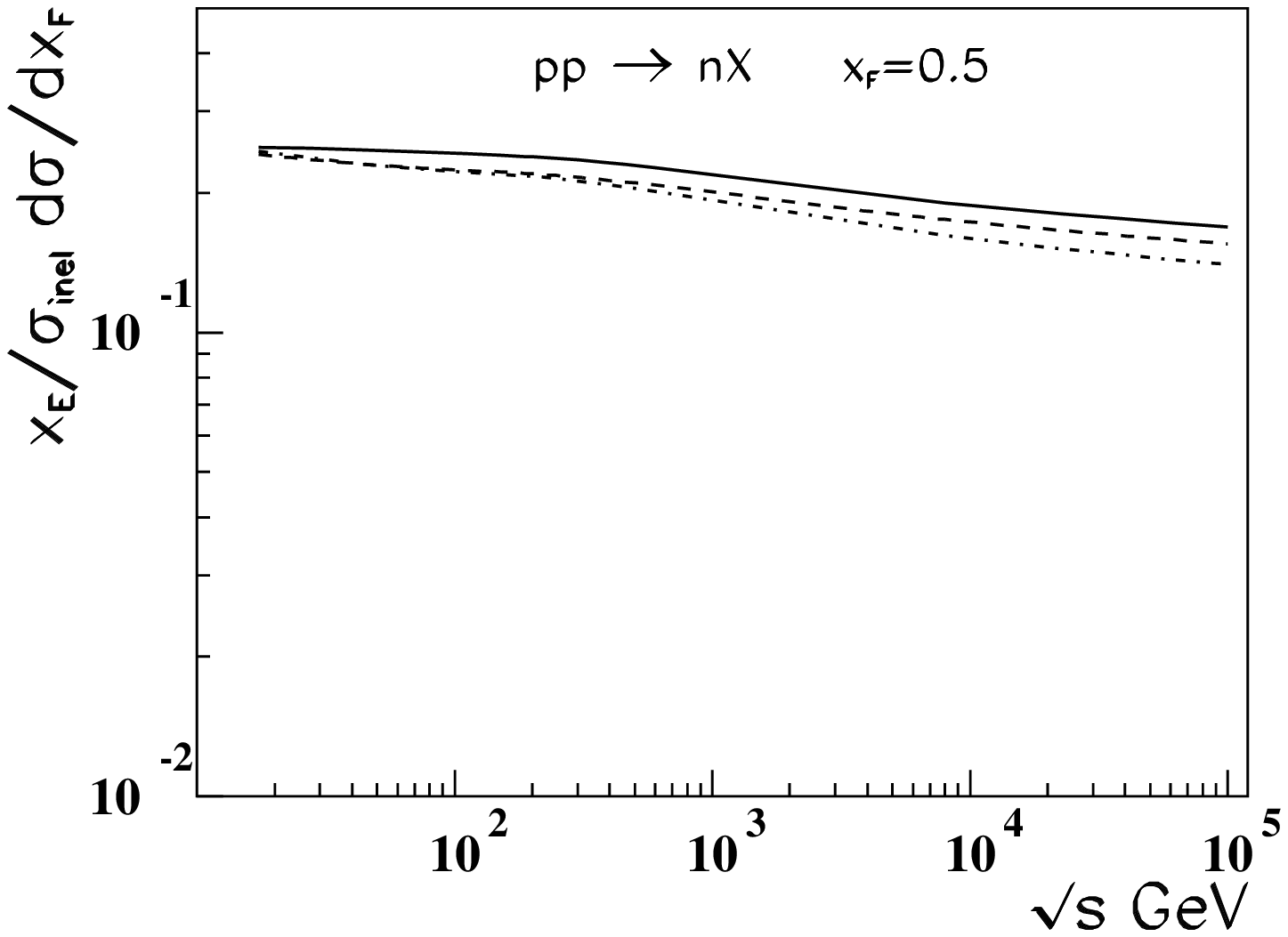}
\vskip -.4cm
\includegraphics[width=.49\hsize]{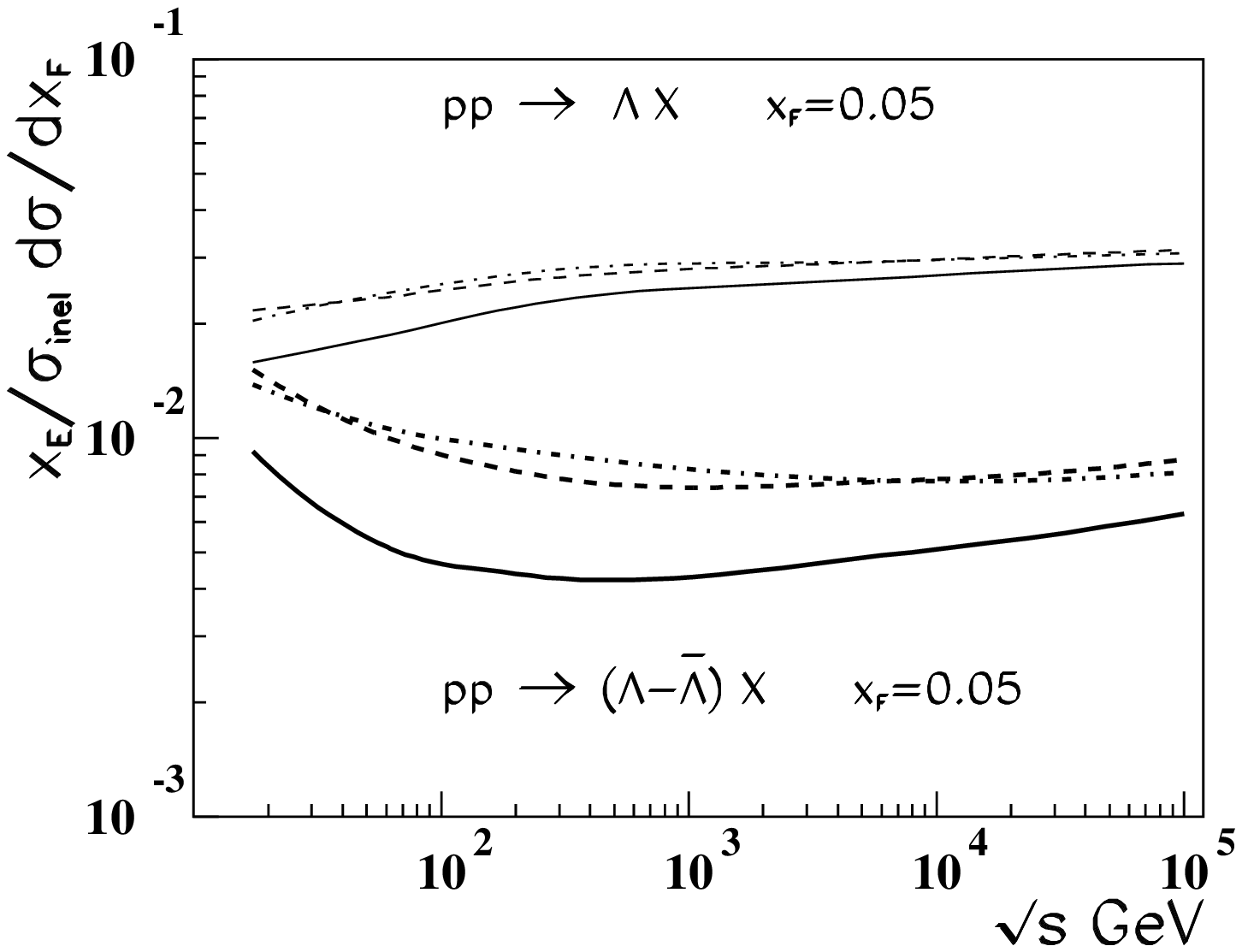}
\includegraphics[width=.49\hsize]{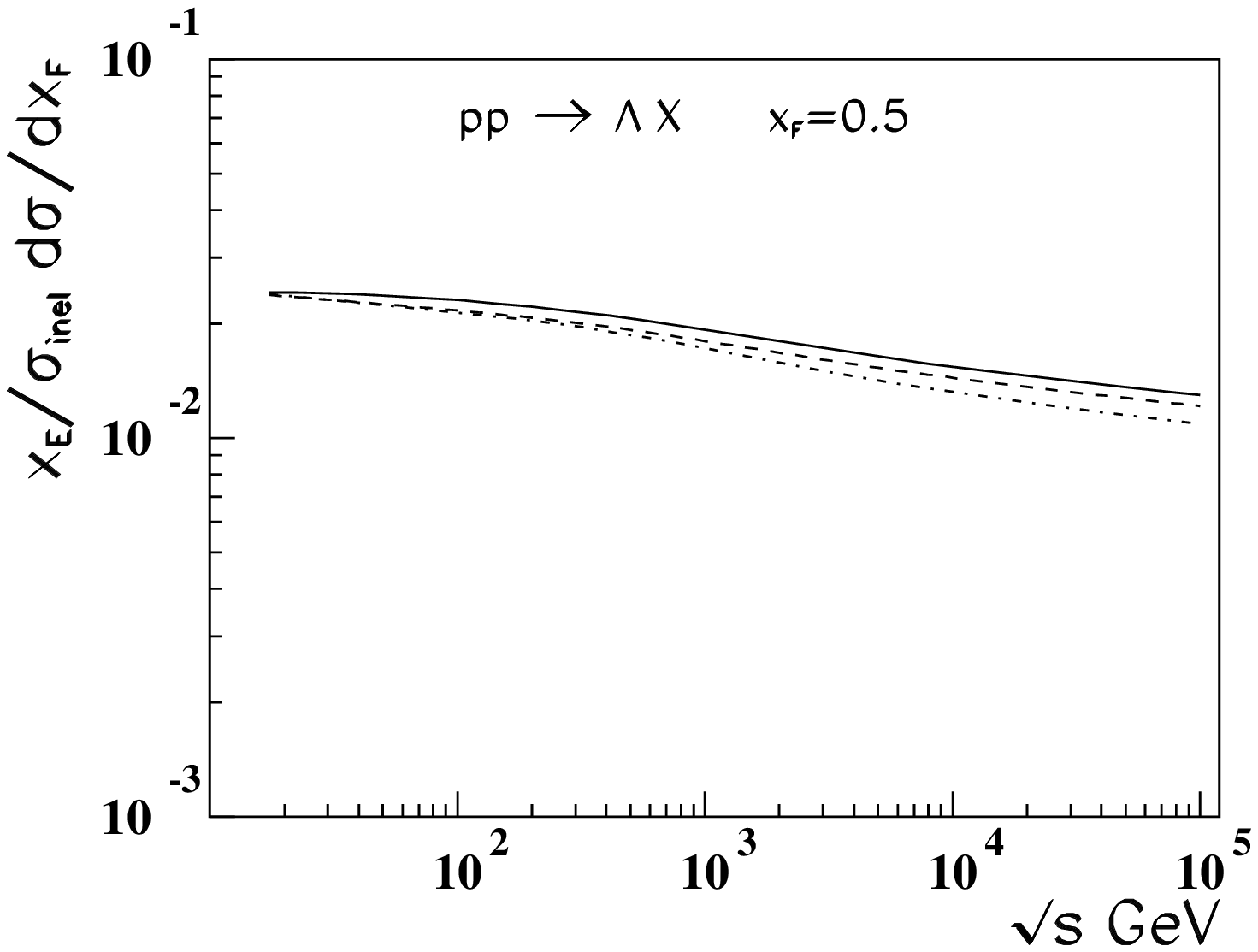}
\vskip -.4cm
\caption{\footnotesize
The QGSM predictions for the spectra of secondary neutrons (top) and
$\Lambda$-hyperons (low) as the functions of energy at fixed values 
of $x_F$. The bold curves (lower sets of curves) on the two left panels
show the spectra of net neutrons and $\Lambda$-hyperons. For the four panel
solid curves correspond to the calculation without SJ contribution, dashed 
curves to the value $\alpha_{SJ} = 0.5$, and dash-dotted curves to the value
$\alpha_{SJ} = 0.9$.}
\end{figure}
For both secondary neutron and $\Lambda$ the spectra increase with
energy at $x_F$ = 0.05 and they decrease at $x_F$ = 0.5. The predicted spectra
of net baryons have a more complicated energy dependence at $x_F$ = 0.05, at
$x_F$ = 0.5 practically coinciding with the total baryon spectra.

Sometimes it is more suitable to compare the spectra in the rapidity variable.
In Fig.~11 we present the spectra of secondary protons, neutrons, and
$\Lambda$ at $\sqrt{s} = 17$ GeV and 8 TeV, as functions of the rapidity
measured from the beam, $y_{beam}-y$, defined in the c.m. frame.
\begin{figure}[htb]
\centering
\vskip -.4cm
\includegraphics[width=.65\hsize]{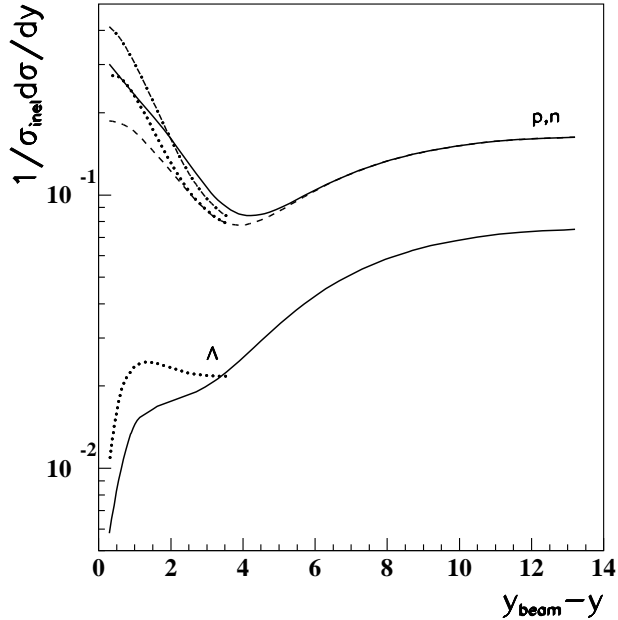}
\vskip -.4cm
\caption{\footnotesize
The QGSM predictions for the spectra of secondary protons, neutrons, and 
$\Lambda$-hyperons produced in $pp$ collisions at $\sqrt{s} = 8$ TeV (solid
and dashed curves), and at $\sqrt{s} = 17$ GeV (dash-dotted and dotted curves
with $\alpha_{SJ} = 0.5$.}
\end{figure}
If Feynman scaling would be preserved, the curves in the left-hand region of
Fig.~11 (i.e. in the beam fragmentation region) should coincide at different
energies, the appearing differences between curves showing the violation of
Feynman scaling. All curves are terminated at the point $y_{cm} = 0$. 

\section{Conclusion}

We present the QGSM predictions for Feynman scaling violation in the spectra 
of leading baryons due both to the increase of the number of cutted Pomerons 
with the energy and to the baryon charge diffusion at large distances in the
rapidity space. The experimental search of the spectra of leading baryons at
LHC should allow to confirm or discard these effects. The possible 
violation of Feynman scaling in the fragmentation region at very high 
energies was discussed in \cite{LHCf1}, based on Monte Carlo calculations. 
In the QGSM hese effects exist, but are, as a rule, not numerically large. 

We have neglected the possibility of interactions between Pomerons (the so-called
enhancement diagrams), since our estimations \cite{MPS2} show that the inclusive
density of secondaries produced in $pp$ collisions at LHC energies is not large
enough for these diagrams to be significant.

Concerning the LHCf project we can note that the multiplicity of 
$\Lambda$ in the fragmentation region should be of the order of 10$-$15\% the
neutron multiplicity, and so it should be accouted for. The production of 
other hyperons should be several orders of magnitude suppressed. 

We have also to note that our results are in reasonable agreement with the
calculations in ref.~\cite{BBKZ}.

A more detailed version of part of the calculations we include in this paper
was already published in ref.~\cite{AMPSh11}.

{\bf Acknowledgements}

We are grateful to \frame{A.B. Kaidalov} for useful discussions
and comments. This paper was supported by Ministerio de Educaci\'on y
Ciencia of Spain under the Spanish Consolider-Ingenio 2010 Programme
CPAN (CSD2007-00042) and project FPA 2005--01963, by Xunta de Galicia, Spain,
and, in part, by grant RFBR 11-02-00120-a and by State Commitee of Science 
of the Republic of Armenia, Grant-11-1C015.




\begin{thebibliography}{**}

\bibitem{Abr} J. Abraham $et~al$., Phys. Rev. Lett. {\bf 104}, 091101 (2010)
and arXiv:1002.0699 [astro-ph.HE].

\bibitem{Abb} R. U. Abbasi  $et~al$., Phys. Rev. Lett. {\bf 104}, 161101 (2010)
and arXiv:0910.4184 [astro-ph.HE].

\bibitem{ELF} E. L. Feinberg, Phys. Report. {\bf 50}, 237 (1972).

\bibitem{Ver} S. N. Vernov  $et~al$., J. Phys. {\bf C~3}, 1601 (1977).

\bibitem{KKh} N. N. Kalmykov and G. B. Khristiansen, JETP Lett. {\bf 37},
294 (1983); Pisma Zh. Eksp. Teor. Fiz. {\bf 37}, 247 (1983).

\bibitem{LHCf1} T. Sako (LHCf Collab.), arXiv:1010.0195 [hep-ex].

\bibitem{LHCf2} O. Adriani  $et~al.$ (LHCf Collab.), arXiv:1012.1490
[hep-ex].

\bibitem{ASS} V. V.~Anisovich, Yu. M. Shabelski, and V. M.~Shekhter,
Nucl. Phys. B~{\bf 133}, 477 (1978).

\bibitem{ABS} V. V.~Anisovich, V. M.~Braun, and Yu. M. Shabelski,
Z. Phys. C~{\bf 27}, 77 (1985).

\bibitem{KTM} A. B. Kaidalov and K.A. Ter-Martirosyan, Sov. J. Nucl. Phys.
{\bf 39}, 979 (1984); Yad. Fiz. {\bf 39}, 1545 (1984); Sov. J. Nucl. Phys.
{\bf 40}, 135 (1984); Yad. Fiz. {\bf 40}, 211 (1984).

\bibitem{Kaid} A. B. Kaidalov, Phys. Atom. Nucl {\bf 66}, 1994 (2003); 
Yad. Fiz. {\bf 66}, 2044 (2003).

\bibitem{KaPi} A. B. Kaidalov and O. I.~Piskounova, Sov. J. Nucl. Phys.
{\bf 41}, 816 (1985); Yad. Fiz. {\bf 41}, 1278 (1985); Z. Phys. {\bf C 30},
145 (1986).

\bibitem{Sh} Yu. M. Shabelski, Sov. J. Nucl. Phys. {\bf 44}, 177 (1986);
Yad. Fiz. {\bf 44}, 186 (1986).

\bibitem{Ans} F. Anselmino, L. Cifarelli, E. Eskut, and Yu.M. Shabelski,
Nouvo Cim. {\bf 105A}, 1371 (1992).

\bibitem{ACKS} G. H.~Arakelyan, A. Capella, A. B.~Kaidalov, and
Yu. M.~Shabelski, Eur. Phys. J. C~{\bf 26}, 81 (2002) and hep-ph/0103337.

\bibitem{KTMS} A. B. Kaidalov, K. A. Ter-Martirosyan, and Yu. M.~Shabelski,
Sov. J. Nucl. Phys. {\bf 43}, 822 (1986); Yad. Fiz. {\bf 43}, 1282 (1986).

\bibitem{Sh1} Yu. M. Shabelski, Z. Phys. {\bf C38}, 569 (1988).

\bibitem{JDDS} J. Dias de Deus and Yu. M. Shabelski, Phys. Atom. Nucl. {\bf 71},
190 (2008).

\bibitem{MPRS} C.~Merino, C.~Pajares, M. M.~Ryzhinskiy, and Yu. M.~Shabelski,
arXiv:1007.3206[hep-ph].

\bibitem{MPS} C. Merino, C. Pajares, and Yu. M.~Shabelski, Eur. Phys. J.
{\bf C71}, 1652 (2011) and arXiv:1105.6026 [hep-ph].

\bibitem{AMPSh11} G. H. Arakelyan, C. Merino, C. Pajares, and Yu. M. Shabelski, arXiv:1107.1615[hep-ph]. 

\bibitem{Sh2} Yu. M. Shabelski, Sov. J. Nucl. Phys. {\bf 45}, 143 (1987); Yad.
Fiz. {\bf 45}, 223 (1987).

\bibitem{Sh3} Yu. M. Shabelski, Z. Phys. {\bf C38}, 569 (1988).

\bibitem{EKS} A. D. Erlykin, N. P. Krutikova, and Yu. M. Shabelski, Sov. J.
Nucl. Phys. {\bf 45}, 668 (1987);Yad. Fiz. {\bf 45}, 1075 (1987); Sov. J.
Nucl. Phys. {\bf 47}, 1057 (1988); Yad. Fiz. {\bf 47}, 1667 (1988).

\bibitem{BS} F. Bopp and Yu. M. Shabelski, Phys. Atom. Nucl. {\bf 68}, 2093
(2005); Yad. Fiz. {\bf 68}, 2155 (2005) and hep-ph/0406158; Eur. Phys. J.
A~{\bf 28}, 237 (2006) and hep-ph/0603193.

\bibitem{AMPS} G. H. Arakelyan, C. Merino, C. Pajares, and Yu. M.~Shabelski,
Eur. Phys. J. {\bf C 54}, 577 (2008) and hep-ph/0709.3174.

\bibitem{MRS1} C. Merino, M. M. Ryzhinskiy, and Yu. M.~Shabelski, Eur. Phys. J.
{\bf C 62}, 491 (2009) and arXiv:0810.1275 [hep-ph].

\bibitem{AKMS} G. H. Arakelyan, A. B. Kaidalov, C. Merino, and Yu. M.~Shabelski,
Phys. Atom. Nucl. {\bf 74}, 426 (2011); Yad. Fiz. {\bf 74}, 447 (2011) 
and arXiv:1004.4074 [hep-ph].

\bibitem{AGK} V. A. Abramovsky, V. N. Gribov, and O. V.~Kancheli, Sov. J.
Nucl. Phys. {\bf 18}, 308 (1974); Yad. Fiz. {\bf 18}, 595 (1973).

\bibitem{Kai} A. B. Kaidalov, Sov. J. Nucl. Phys. {\bf 45}, 902 (1987);
Yad. Fiz. {\bf 43}, 1282 (1986).

\bibitem{Kar3} K. A. Ter-Martirosyan, Phys. Lett. {\bf B44}, 377 (1973).

\bibitem{PAO} P. Abreu  $et~al$., {\it In Proceedings of the 32nd International 
Cosmic Ray Conference}, Beijing, China, 2011; arXiv1107.4809[astro-ph.HE].

\bibitem{Artru} X. Artru, Nucl. Phys. B {\bf85}, 442 (1975).

\bibitem{IOT} M. Imachi, S. Otsuki, and F.~Toyoda, Prog. Theor. Phys.
{\bf 52}, 346 (1974); {\bf 54}, 280 (1976); {\bf 55}, 551 (1976).

\bibitem{RV} G. C. Rossi and G.~Veneziano, Nucl. Phys. B~{\bf 123}, 507 (1977).

\bibitem{Khar} D. Kharzeev, Phys. Lett. B~{\bf 378}, 238 (1996).

\bibitem{latt} V. G. Bornyanov  $et~al$, Uspekhi Fiz. Nauk. {\bf 174},
19 (2004).

\bibitem{AMS} G. H. Arakelyan, C. Merino, and Yu. M.~Shabelski, Yad. Fiz.
{\bf 69}, 911 (2006) and hep-ph/0505100; 
Yad. Fiz. {\bf 70}, 1146 (2007); Phys. Atom. Nucl. {\bf 70}, 1110(2007) 
and hep-ph/0604103; Eur. Phys. J. {\bf A31}, 519 (2007) and hep-ph/0610264.

\bibitem{Olga} O. I. Piskounova, Phys. Atom. Nucl. {\bf 70}, 1107 (2007);
Yad. Fiz. {\bf 70}, 1143 (2007) and hep-ph/0604157.

\bibitem{MRS} C. Merino, M. M. Ryzhinskiy, and Yu. M.~Shabelski,
{\it Proceedings of the XLIII PNPI Winter School on Nuclear and Particle
Physics (PNPI-2009)}, Repino, St.Petersburg, Russia, February
$24{\rm th}$-March $1{\rm st}$, 2009, pages 156-185, and arXiv:0906.2659 [hep-ph].

\bibitem{VGW} S. E. Vance, M. Gyulassy, and X-N. Wang, Phys. Lett. B~{\bf 443}, 45 (1998).

\bibitem{ait} E. M. Aitala {\em et al.} (E769 Collab.), hep-ex/0009016;
Phys. Lett. B~{\bf469}, 9 (2000).

\bibitem{KP1} B. Z. Kopeliovich and B.~Povh, Z. Phys. C~{\bf 75}, 693 (1997);
Phys. Lett. B~{\bf 446}, 321 (1999).

\bibitem{AnSh} V. V.~Anisovich and V. M.~Shekhter, Nucl. Phys. B~{\bf 55}, 455 (1973).

\bibitem{CS} A. Capella and C. A. Salgado, Phys. Rev. C~{\bf 60}, 054906 (1999).

\bibitem{ALICE} K. Aamodt $et~al.$ (ALICE Collab.),
Phys. Rev. Lett. {\bf 105}, 072002 (2010) and arXiv:1006.5432 [hep-ex].

\bibitem{LHCb} F. Dettori $et~al.$ (LHCb Collab.), Nucl. Phys. B 
(Proc. Suppl.) {\bf 206-207}, 151 (2010) and arXiv:1009.1221 [hep-ex].

\bibitem{MPS3} C. Merino, C. Pajares, and Yu. M.~Shabelski, arXiv:1105.6026 [hep-ph].

\bibitem{Conf} C. Merino, C. Pajares, M. M. Ryzhinskiy, and Yu. M.~Shabelski, Proceedings of 
the International Conference on Hadron Structure and QCD-HSQCD 2010, edited by V. T. Kim and L. N. Lipatov,
PNPI Publishing Department, Gatchina, Russia, 5-9 July 2010, pages 75-82.

\bibitem{na49p} T. Anticic  $et~al.$ (NA49 Collab.), Eur. Phys. J. {\bf C65},
9 (2010) and arXiv:0904.2708 [hep-ex].

\bibitem{Bren} A. E. Brenner, Phys. Rev. {\bf D26}, 1497 (1982).

\bibitem{MPS2} C. Merino, C. Pajares, and Yu. M.~Shabelski, Eur. Phys. J.
{\bf C59}, 691 (2009) and arXiv:0802.2195 [hep-ph].

\bibitem{BBKZ} C. J. Bleibel  $et~al.$, arXiv:1011.2703 [hep-ph].

\end{thebibliography}
\end{document}